# Properties of the Abelian Projection Fields in $SU(N)$ Lattice Gluodynamics

M.I. Polikarpov

*ITEP, Moscow, 117259, Russia*

e-mail: polykarp@vxdsyc.desy.de

Ken Yee

*Dept. of Physics and Astronomy, L.S.U.*
*Baton Rouge, Louisiana  70803-4001, USA*

e-mail: kyee@rouge.phys.lsu.edu

### Abstract

't Hooft's abelian projection of $SU(N)$ gauge theory yields $N$ mutually constrained, compact abelian fields which are permutationally equivalent. We formulate the notion of "species permutation" symmetry of the $N$ abelian projection fields and discuss its consequences for cross-species correlators. We show that at large $N$ cross-species interactions are $\frac{1}{N}$ suppressed relative to same-species interactions. Numerical simulations at $N = 3$ support our symmetry arguments and reveal the existence of inter-species interactions of size $\mathcal{O}(\frac{1}{N-1})$ as analytically predicted.

# 1  Monopole and Photon Species in $SU(3)$

Since 't Hooft's invention of the abelian projection approach to isolating $U(1)$ monopoles in color $SU(N)$ gauge theories [1], numerical lattice simulations of $SU(2)$ [2] and $SU(3)$ [3] have revealed that these abelian $SU(N)$ monopoles are especially interesting in maximal abelian gauge(MAG). Gauge fixing to MAG reduces the local gauge symmetry from $SU(N)$ to $[U(1)]^{N-1}$. Imagining that the dynamical degrees of freedom associated with coset space $SU(N)/[U(1)]^{N-1}$ are integrated out, one can regard what remains as an effective $[U(1)]^{N-1}$ gauge theory whose dynamical variables are $N$ abelian link angles $\theta_i(x,\mu)$ constrained by one sum rule (to be given). Indices $i,j,k,l \in \{1, \cdots, N\}$ label the color $N$ abelian "species" and $N \geq 2$ is assumed in this paper.

A direction recently pursued by several groups [4]-[10] is to try and characterize confinement properties of the QCD vacuum with an effective $[U(1)]^{N-1}$ action, which to date is unknown even for $N = 2$, the simplest case, and $N = 3$, the *physical* case [10]. Early simulations [2, 3] established that, viewed independently, each species of this $[U(1)]^{N-1}$ gauge theory is crudely reminiscent of compact QED [11], that is, the monopoles are dense and kinetic in the confinement region and dilute and static in the deconfined regime.

On the other hand, if the abelian projection picture of confinement holds all the way down to the continuum limit, the effective $[U(1)]^{N-1}$ gauge theory cannot be simply $N$ constrained copies of ordinary compact QED in this limit. This is clear because compact QED in $D = 3+1$ dimensions confines only in a noncritical region with, for example, no continuous rotational sym-metry while continuum QCD is Lorentz invariant. Indeed, while numerical simulations show that compact QED within the confining phase $\beta < \beta_c$ is an extreme Type II superconductor [12], abelian projected $SU(2)$ QCD seems to fall on the interface between Type I and Type II superconductivity [6].

In this paper, to lay the groundwork for progress in the $N \geq 3$ cases we consider the following questions:

- How are the $N$ abelian projection $U(1)$ fields related to each other? How do the different $U(1)$ species interact?

- How are the $N$ monopole currents related to the $N$ abelian electric fields?

In Section 2 we report on numerical computations in $SU(3)$ of some fundamental $[U(1)]^2$ local correlation functions. In Section 3 we explain, based on just broad sum rule and species symmetry considerations, the observed numerical relationship between these correlation functions. We show that in the $N \to \infty$ limit the different $U(1)$ species decouple linearly with decreasing $\frac{1}{N}$. Appendix A describes the implementation of the $SU(N)$ abelian projection on the lattice. Appendix B gives the derivations of mathematical formulas cited in Section 3.

## 2  $U(1)$ and $SU(3)$ Numerical Results

The link angles $\theta_i(x,\mu)$ form $N$ constrained plaquette angles $\Theta_i(P) \equiv \Theta_i(x,\mu\nu)$ where $P$ denotes a plaquette in the lattice. In turn, the plaquette angles form $N$ constrained monopole currents $k_i({}^*x,\mu)$. Definitions of $\theta_i$, $\Theta_i$, and $k_i$ in terms of the original $SU(N)$ links are given in Appendix A and are entirely consistent with Refs. [3, 13].

In this Section we summarize $SU(3)$ results for local correlation functions involving $k_i$, its curl $\nabla \times k_i$, and abelian electric field[1]

$$E_i(x, \mu) \equiv \sin \Theta_i(x, \mu 0) \qquad \mu \in \{1, 2, 3\}. \tag{1}$$

We are interested in not only same-species but also cross-species correlations. To this end, let $u_i$ and $v_i$ be $i^{\text{th}}$-species vector operators such as $k_i$, $\nabla \times k_i$ or $E_i$. Define normalized parallel correlator

$$[u_i, v_j]_{\|} \equiv \frac{(u_i, v_j)_{\|}}{\sqrt{(u_i, u_i)_{\|}(v_j, v_j)_{\|}}} \tag{2}$$

where

$$(u_i, v_j)_{\|} \equiv \frac{1}{4 \text{ VOL}} \sum_{x,\mu} \langle u_i(x, \mu) v_j(x, \mu) \rangle. \tag{3}$$

$[u_i, v_j]_{\|}$ is a dimensionless measure of the virtual directional correlation of $u_i$ with $v_j$. $\langle \cdots \rangle$ refers to the abelian-projected QCD expectation value.

Our interest in the aforementioned operators and correlation functions is motivated by properties of compact QED described in Ref. [12]. In compact QED monopole currents, playing the role of cooper pairs in superconductors, circulate around the electric flux tube between a widely separated static quark-antiquark($q\bar{q}$) pair. The circulating monopole currents are responsible for confinement, that is, for "squeezing" the electric $q\bar{q}$ flux into an Abrikosov tube. Since *virtual* electromagnetic flux exists even in the absence of external sources, the curl of $k$ should be nontrivially correlated to electric field $E$—at least within the $\beta < \beta_c$ phase. Indeed, as depicted in Figure 1

$$\|[\nabla \times k, E]_{\|}\| \;\gg\; \|[k, E]_{\|}\| \;\sim\; 0 \tag{4}$$

for $\beta < \beta_c$. Interpretationally, the expectation value of $\nabla \times k$ is locally parallel to $-E$ whereas $k$ is relatively uncorrelated to $E$ in compact QED.

---

[1] $\mu \in \{0, 1, 2, 3\}$ for $\theta_i$ and $k_i$, and $\mu \in \{1, 2, 3\}$ for $\nabla \times k_i$ and $E_i$. Note that "$\mu 0$" in Eq. (1) refers to the two plaquette indices.

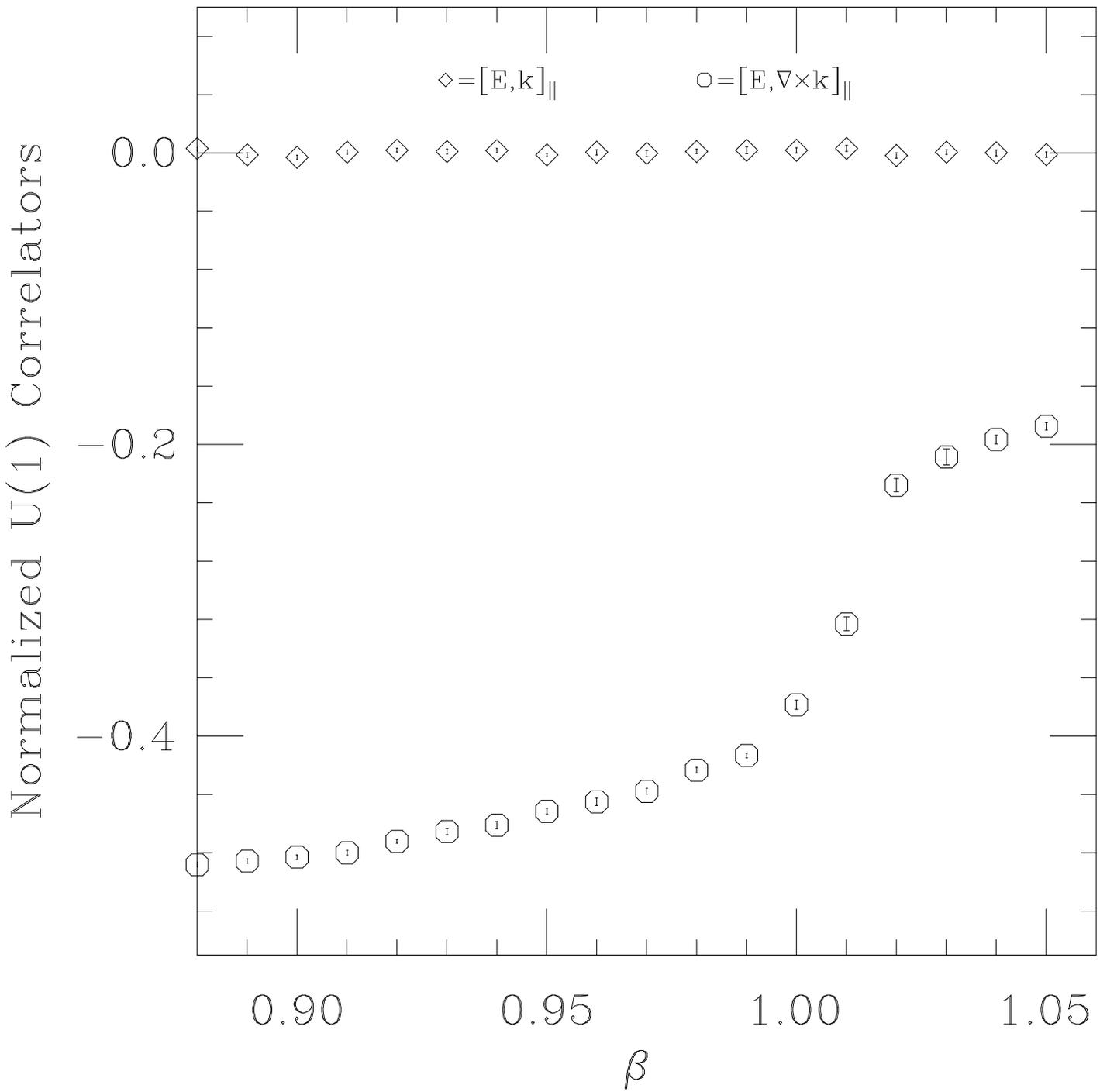

Figure 1: $[E,k]_{\|}$ and $[E,\nabla\times k]_{\|}$ in compact QED as a function of $\beta$ on $8^3 \times 16$ lattices. At each $\beta$ the first configuration is thermalized from a fresh cold start by 2000 24-hit Metropolis/pseudoheatbath sweeps. Each configuration thereafter is separated by 15 sweeps, and 10 configurations are used for each $\beta$. Error bars are jackknife errors.

| $i$ | $j$ | $u$ | $v$ | $16^3 \times 24$ | $24^3 \times 40$ |
|---|---|---|---|---|---|
| 1 | 1 | $k$ | $k$ | 1.0 | 1.0 |
| 1 | 2 | $k$ | $k$ | -.50(.002) | -.51(.004) |
| 1 | 3 | $k$ | $k$ | -.50(.001) | -.49(.003) |
| 1 | 1 | $E$ | $E$ | 1.0 | 1.0 |
| 1 | 2 | $E$ | $E$ | -.44(.001) | -.48(.001) |
| 1 | 3 | $E$ | $E$ | -.44(.001) | -.48(.001) |
| 1 | 1 | $k$ | $E$ | .0002(.0004) | .00014(.00004) |
| 1 | 2 | $k$ | $E$ | .0006(.001) | -.00016(.0003) |
| 1 | 3 | $k$ | $E$ | -.0007(.001) | .00003(.0003) |
| 1 | 1 | $E$ | $\nabla \times k$ | -.28(.001) | -.095(.001) |
| 1 | 2 | $E$ | $\nabla \times k$ | .14(.001) | .048(.001) |
| 1 | 3 | $E$ | $\nabla \times k$ | .14(.001) | .047(.001) |

Table 1: Parallel correlator $[u_i, v_j]_\parallel$ between different species of monopole current $k$, its curl, and electric field $E$ on $\beta = 5.7$, $16^3 \times 24$ and $\beta = 6.0$, $24^3 \times 40$ $SU(3)$ lattices. The normalized expectation value $[u_i, v_j]_\parallel$ is defined in the text.

This feature is also realized in abelian-projected QCD. Table 1 lists the $SU(3)$ correlators on, respectively, 16 (decorrelated) configurations of $16^3 \times 24$ lattices at $\beta = 5.7$ and 8 (also decorrelated) configurations of $24^3 \times 40$ lattices at $\beta = 6.0$ [14]. Both lattices are fixed to MAG by the usual methods [3]. The Table shows that, just like in the confining phase of compact QED, $E_i$ is much more correlated to $\nabla \times k_i$ than to $k_i$.

We observe in Table 1 that $[E_1, \nabla \times k_1]_\parallel$, which is (naively) dimensionless and nonzero by many jackknife errors on both lattices, violates scaling rather dramatically between $\beta = 5.7$ and $\beta = 6.0$.

Naively one might expect that $-E_j$ for $j \neq i$ is also positively correlated to $\nabla \times k_i$, e.g., that all monopole species circulate in the same direction around virtual $-E_j$ flux. On the contrary Table 1 reveals that $-E_j$ is *anti*-correlated to $\nabla \times k_i$. Such a situation arises due to the general properties of the inter-species interaction discussed in the next Section.

## 3  Cross-Species Interactions

Let calligraphic letters $\{\mathcal{A}_i = \mathcal{A}_i(\theta_i),\ \mathcal{B}_i = \mathcal{B}_i(\theta_i),\ \cdots\}$ denote possibly non-local, possibly composite operators comprised exclusively of $i^{\text{th}}$-species link angles. Let Roman letters $\{c_i, d_i, \cdots\}$ denote special $i^{\text{th}}$-species operators satisfying the "sum rule" constraint

$$\sum_{i=1}^{N} c_i = 0. \tag{5}$$

As defined in Appendix A, $\theta_i$, $\Theta_i$, $k_i$ and $\nabla \times k_i$ all satisfy (5). In the next Section we prove that

$$\langle \mathcal{A}_j\ c_k \rangle = -\left(\frac{1}{N-1}\right)\langle \mathcal{A}_i\ c_i \rangle \qquad j \neq k. \tag{6}$$

In this Section, we describe the physical consequences of (6) which, as the Reader can check, is consistent with the data in Table 1 and explains why, for example, that $[k_1, k_2]_\parallel = -\frac{1}{2}[k_1, k_1]_\parallel$ for $N = 3$. Moreover, since

$$\sum_{i=1}^{N} E_i = \lim_{a \to 0} \sum_{i=1}^{N} \sin(a^2 \Theta_i) = a^2 \sum_{i=1}^{N} \Theta_i = 0 \tag{7}$$

where $a$ is the lattice spacing,[2] $E_i$ approximately satisfies (5) in the continuum limit. Indeed, as shown in Table 1 $[E_i, E_j]_\parallel$ approaches (6) as $\beta$ increases from 5.7 to 6.0.

What does (6) tell us about the confining vacuum from the abelian-projection vantage point? Consider [4, 12]

$$\overline{c}_i^j \equiv -i\,\frac{\langle W_j\ c_i \rangle}{\langle W_j \rangle} \tag{8}$$

where $W_j$ is the $j^{\text{th}}$-species time-like abelian Wilson loop which we take to be suitably much larger than the abelian flux tube width. $\overline{c}_i^j$ is the expectation value of operator $c_i$ in the background electric field created by a widely-separated static $(q\bar{q})_j$ pair.

---

[2]By identification the effective $[U(1)]^{N-1}$ lattice spacing is regarded to be equal to the original $SU(N)$ QCD lattice spacing.

Eq. (6) implies that

$$\overline{c}_i^j = -\Big(\frac{1}{N-1}\Big)\overline{c}_j^j \quad i \neq j \qquad (9)$$

where there is no sum over repeated $j$'s on the RHS. A physical interpretation emerges if, for example, we set $c_i = E_i$. Then (9) implies that the $i^{\text{th}}$ effective electric field $\overline{E}_i^j$ between a $j^{\text{th}}$-species $q\bar{q}$ pair points in the opposite direction of $\overline{E}_j^j$ and is suppressed in magnitude relative to $\overline{E}_j^j$ by $\frac{1}{N-1}$. The same analysis also applies to $\overline{\nabla \times k}_i^j$.

Note that since $\langle \cdots \rangle$ in numerical simulations is given in terms of an importance sampling sum, it is a manifestly positive definite dot product and the Cauchy-Schwartz inequality $|\langle \mathcal{A}_i \mathcal{B}_i \rangle| \leq |\langle \mathcal{A}_i \mathcal{A}_i \rangle|^{\frac{1}{2}} |\langle \mathcal{B}_i \mathcal{B}_i \rangle|^{\frac{1}{2}}$ applies. Together with (6), Cauchy-Schwartz implies that

$$|\langle \mathcal{A}_i \; c_j \rangle| \leq \Big(\frac{1}{N-1}\Big)|\langle \mathcal{A}_i \mathcal{A}_i \rangle|^{\frac{1}{2}} \; |\langle c_j c_j \rangle|^{\frac{1}{2}} \quad i \neq j. \qquad (10)$$

Therefore, as long as $\mathcal{A}_i$ and $c_j$ have finite norm, they become uniformly uncorrelated *linearly* with $\frac{1}{N}$ as $N \to \infty$ limit. This is independent of the spatial shapes of and separation between $\mathcal{A}_i$ to $c_j$.

An interesting consequence of Eq. (10) is that in the $N \to \infty$ limit only same-species correlations survive. Therefore, if the same mechanism which drives $N \to \infty$ confinement drives finite $N$ confinement, *cross species interactions cannot be responsible for confinement.* Nonetheless since the world has $N = 3$ colors, the cross-species interactions may be potentially phenomenologically significant.

## 4 Acknowledgments

MIP thanks the LSU Physics Department for its hospitality. We have benefitted from very many and useful discussions with Dick Haymaker, especially regarding $\nabla \times k$. KY also thanks Lai Him Chan and Dana Browne for discussions, Gerrit Schierholz for Ref. [13], and Claude Bernard and Amarjit Soni for the use of their NERSC(Livermore) account and $SU(3)$ gauge configurations. Analysis was done on the NERSC grant of KY. MIP has been partially supported by a grant of the American Physical Society. KY is supported by DOE grant DE-FG05-91ER40617.

# Appendix A  MAG and the Lattice Abelian Projection

First, a cautionary note about numerical maximal abelian gauge fixing. Fixing to MAG involves picking local gauge transformations $V(x)$ so that an operator "$X(x)$" is diagonalized. Since inequivalent orderings of the eigenvalues of $X(x)$ are associated with different $V(x)$'s, this condition has inherent gauge fixing ambiguities. In 't Hooft's original paper [1, 2] he suggested ordering the eigenvalues of $X(x)$ by size to eliminate the ambiguities. However, as far as we know, none of the existing numerical MAG algorithms (including our own) implement this auxiliary condition and such an ordering ambiguity is always implicitly present in numerical simulations.

Our lattice abelian projection scheme is the same as that in Refs. [3, 13] and we state the construction here for completeness. The abelian projection in MAG(or any other appropriate gauge of interest) is given by

$$\theta_i(x,\mu) \equiv \arg(\widetilde{U}_{ii}(x,\mu)) - \frac{1}{N}\phi(x,\mu) \quad \in \left(\frac{N+1}{N}\right) \times [-\pi,\pi], \quad (A.1)$$

$$\phi(x,\mu) \equiv \Big[\sum_{i=1}^{N} \arg(\widetilde{U}_{ii}(x,\mu))\Big] \bmod 2\pi \quad \in [-\pi,\pi) \quad (A.2)$$

where $\widetilde{U}(x,\mu)$ are $SU(N)$ links fixed to MAG. Definition (A.1) implies

$$\sum_{i=1}^{N} \theta_i(x,\mu) = 0, \quad (A.3)$$

a constraint required because there are only $N-1$ independent abelian fields.

The plaquette angles $\Theta_i(P)$ are given by the oriented sum of link angles $\theta_i$ around $P$ with the additional condition that the $2\pi$ ambiguity in the plaquette phase is fixed so that

$$\sum_{i=1}^{N} \Theta_i(P) = 0. \quad (A.4)$$

This is achieved by introducing the intermediate variable

$$\tilde{\Theta}_i(P) = \Theta_i(P) \bmod 2\pi \quad \in [-\pi, \pi). \tag{A.5}$$

By Eqs. (A.3) and (A.5), $\sum_{i=1}^{N} \tilde{\Theta}_i(P) = 2\pi n$ where, specializing now to $N = 3$, integer $n$ can be $\{0, \pm 1\}$. If $n = +1$, the plaquette phases are chosen so that

$$\Theta_i = \begin{cases} \tilde{\Theta}_i - 2\pi & \text{if } \tilde{\Theta}_i = \max(\tilde{\Theta}_1, \tilde{\Theta}_2, \tilde{\Theta}_3); \\ \tilde{\Theta}_i & \text{otherwise.} \end{cases} \tag{A.6}$$

If $n = -1$,

$$\Theta_i = \begin{cases} \tilde{\Theta}_i + 2\pi & \text{if } \tilde{\Theta}_i = \min(\tilde{\Theta}_1, \tilde{\Theta}_2, \tilde{\Theta}_3); \\ \tilde{\Theta}_i & \text{otherwise.} \end{cases} \tag{A.7}$$

The $\Theta_i$ defined in this way obey (A.4). Note that one of the three $\Theta_i$ angles at each plaquette may lie outside $[-\pi, \pi)$. The $N > 3$ cases would be handled analogously, always keeping in mind the preservation of species permutation symmetry.

The monopole current is given by

$$k_i(^*x, \mu) \equiv \frac{1}{4\pi} \sum_{P \in C} \Theta_i(P) = \{0, \pm\frac{1}{2}, \pm 1\} \tag{A.8}$$

where $C$ is the cube at dual lattice site $^*x$ orthogonal to $\mu$. Plaquette constraint (A.4) implies

$$\sum_{i=1}^{N} k_i(^*x, \mu) = 0. \tag{A.9}$$

This seemingly contrived constraint on the monopole currents can be understood on a deeper level to be a required feature of $SU(N)$ monopole singularities revealed through gauge-fixing [1, 2].

# Appendix B  Species Permutation Symmetry

The $N$ constrained $U(1)$ abelian projection fields has a species permutation symmetry which we state as follows:

**Every species is equivalent to every other species and, for $i \neq j$ and $i \neq l$, the relationship of species $i$ to $j$ is the same as $i$ to $l$.**

In this Appendix we demonstrate how (6) and its extensions follow from Eq. (5) and species permutation symmetry.

Species permutation implies that

$$\langle \mathcal{A}_i \mathcal{B}_i \rangle = \langle \mathcal{A}_j \mathcal{B}_j \rangle, \tag{B.1}$$

$$\langle \mathcal{A}_i \mathcal{B}_j \rangle = \langle \mathcal{A}_i \mathcal{B}_k \rangle, \quad j \neq i, \; k \neq i. \tag{B.2}$$

We emphasize that there is *no implicit summation* over repeated species indices in Eq. (B.1) or in any equation in this paper.

Sum rule (5) and species permutation symmetry implies that

$$\langle c_i \rangle = -\sum_{j \neq i} \langle c_j \rangle = -(N-1) \langle c_j \rangle \tag{B.3}$$

which in turn implies that for $N \geq 2$

$$\langle c_i \rangle = 0. \tag{B.4}$$

Straightforward generalizations of such manipulations yield Eq. (6) and, if $N \geq 3$,

$$\langle \mathcal{A}_i \mathcal{B}_j \; c_k \rangle = -\Big(\frac{1}{N-2}\Big)\big\{\langle \mathcal{A}_i \mathcal{B}_j \; c_j \rangle + \langle \mathcal{A}_i \mathcal{B}_j \; c_i \rangle\big\} \quad k \neq i \neq j \neq k. \tag{B.5}$$

The Reader is invited to derive other such relations involving more complicated correlators at larger $N$.